\documentclass[fleqn]{2023SCGE}
\begin{document}

\ensubject{subject}

%%%%%%%%%%%%%%%%%%%%%%%%%%%%%%%%%%
%%% Authors do not modify the information below
%Letter to the Editor??Article%??????
\ArticleType{Article}%??Article
\SpecialTopic{}
\Year{2026}
\Month{February}
\Vol{69}
\No{2}
\DOI{10.1007/s11433-025-2798-5}
\ArtNo{220411}
\ReceiveDate{July 18, 2025}
\AcceptDate{August 29, 2025}
\OnlineDate{January 4, 2026}
%%%%%%%%%%%%%%%%%%%%%%%%%%%%%%%%%%%

\title{Time delay interferometry with minimal null frequencies and shortened time span}{Time delay interferometry with minimal null frequencies and shortened time span}

\author[1,2,3]{Gang Wang}{gwanggw@gmail.com, gwang@nbu.edu.cn}

\AuthorMark{Wang G}
\AuthorCitation{Wang G}

\address[1]{Institute of Fundamental Physics and Quantum Technology, Ningbo University, Ningbo, 315211, China}
\address[2]{Department of Physics, School of Physical Science and Technology, Ningbo University, Ningbo, 315211, China}
\address[3]{Shanghai Astronomical Observatory, Chinese Academy of Sciences, Shanghai 200030, China}

\abstract{
Time-delay interferometry (TDI) is essential for suppressing laser frequency noise in space-based gravitational wave (GW) observatories such as LISA. However, current second-generation TDI schemes often exhibit undesirable null frequencies and require long delay spans, which can impair data analysis performance. In this work, we introduce an alternative TDI configuration—PD4L—designed to minimize null frequencies and operate with a shorter effective time span. Constructed by synthesizing two distinct first-generation TDI schemes, PD4L achieves a delay span of 4$L$ (where $L$ is the arm length), half that of the standard Michelson and hybrid Relay configurations. We assess PD4L's performance by evaluating the spectral stability of instrumental noise via arm-length derivatives, simulating chirping GW signals from coalescing massive black hole binaries, and comparing waveform responses. Parameter estimation is performed in the frequency domain, and noise characterization is examined under realistic orbital dynamics. As demonstrated by the comparisons, the compact structure of PD4L offers several advantages: (1) reduced data margins at segment boundaries, (2) mitigated aliasing effects in the high-frequency regime, and (3) shortened signal tails arising from extended delay spans. Additionally, PD4L’s null channels exhibit the same minimal null frequencies as its science channels, while maintaining greater spectral stability than other null streams. Overall, PD4L improves parameter estimation accuracy at high frequencies and supports reliable noise characterization over observation periods of up to four months. These results highlight PD4L as a compact and effective alternative for future TDI implementations, especially in high-frequency GW data analysis for LISA-like missions.
}


\maketitle

\begin{multicols}{2}
\section{Introduction}

Time delay interferometry (TDI) was developed to suppress laser frequency noise and achieve the targeted sensitivity for space-borne interferometers \cite{1997SPIE.3116..105N,1999ApJ...527..814A,2000PhRvD..62d2002E}. It is essential for space missions such as LISA \cite{2017arXiv170200786A,Colpi:2024xhw}, TAIJI \cite{Hu:2017mde}, TianQian \cite{TianQin:2015yph} to detect gravitational waves (GW) in the millihertz band. 
\Authorfootnote
TDI is to appropriately delay and combine interferometric laser links to construct equivalent equal-arm interferometry. The first-generation TDI was designed to cancel laser noise in a static unequal-arm case \cite{1999ApJ...527..814A}. However, due to orbital dynamics, second-generation TDI is required to eliminate the effects of relative spacecraft (S/C) motions \cite{Tinto:2003vj,Shaddock:2003dj}.

Among second-generation TDI configurations, the Michelson has been widely used as a benchmark for noise suppressions and data analysis. However, as shown in our earlier investigations \cite{Wang:2024alm,Wang:2024hgv}, its performance degrades in realistic dynamic unequal-arm scenario. To address this, we proposed an alternative TDI scheme, the hybrid Relay, which offers enhanced data analysis capabilities by minimizing null frequencies and improving noise spectral stability. Nevertheless, the hybrid Relay, as the Michelson, requires delays up to $7L$ (where $L$ is the light-travel time between S/C), resulting in a total time span of $8L$. We notice the TDI with longer delay time could introduces disadvantages: larger data margins are generated at segment boundaries, leading to data loss, and longer-span TDI operations traverse more of the waveform's evolution, making them more susceptible to frequency aliasing. 

To mitigate these issues, a more compact TDI configuration with a shorter time span is desirable. We examined a series of alternative second-generation TDI configurations developed in 2011 \cite{Wang:2011}, aiming to identify an optimal scheme with minimized null frequencies and reduced time span.
Among these, we identified PD4L as a promising configuration and implemented it for GW signal analysis for the first time in this work. This configuration was also identified using a different approach in \cite{Muratore:2021uqj}. PD4L features a time span of $4L$ (with a maximum delay of $3L$) and exhibits minimal null frequencies in both ordinary and optimal channels. Compared to the hybrid Relay and Michelson schemes, which span $8L$, PD4L reduces the data combination window by half. 
We conduct a multi-step evaluation of PD4L's performance. First, we analyze the stability of its noise spectra by calculating their derivatives with respect to arm-length variations. Although PD4L's science channels exhibit slightly less robustness than those of the hybrid Relay, it remains sufficiently stable for analysis. An notable advantage is that its null stream exhibits superior spectral stability compared to the $C^{12}_3$ channel proposed in \cite{Hartwig:2021mzw}, making it the most stable null stream among those we investigated. 
We evaluate the sky-averaged GW response and sensitivity. While PD4L observables have weaker responses than that of hybrid Relay and Michelson channels at lower frequencies, their optimal-channel sensitivities remain effectively identical.

To further validate PD4L's performance, we simulate chirping GW signals from coalescing massive binary black holes (MBBHs) and compute the response across Michelson, hybrid Relay, and PD4L schemes. We find that TDI channels with shorter time span produce smoother time-domain waveforms. We also perform parameter inference on the simulated signals in the frequency-domain. The results suggest that PD4L outperforms the hybrid Relay for the high-frequency signal and achieves comparable performance at low frequencies. Finally, we assess PD4L's capability in noise characterization over different data durations, showing that it can accurately determine the noise parameters for data duration up to four months.

This paper is organized as follows: In Section \ref{sec:tdi}, we introduce the second-generation TDI configurations considered in this study. 
Section \ref{sec:noise} analyzes the noise spectra of the TDI observables and their derivatives to arm length variations. In Section \ref{sec:gw_simulation}, we compute the GW responses, evaluate sky-averaged sensitivities, and compare time-domain waveforms for a rapidly chirping GW signal across different TDI observables. 
Section \ref{sec:parameters_inference} presents the parameter inference of chirping signals and assesses noise characterizations with different data durations. 
Finally, a brief conclusion and discussion are given in Section \ref{sec:conclusions}.
(Throughout this work, we set $G=c=1$, unless otherwise stated.)

\section{Time delay interferometry} \label{sec:tdi}

\begin{figure}[H]
\centering
\includegraphics[width=0.4\textwidth]{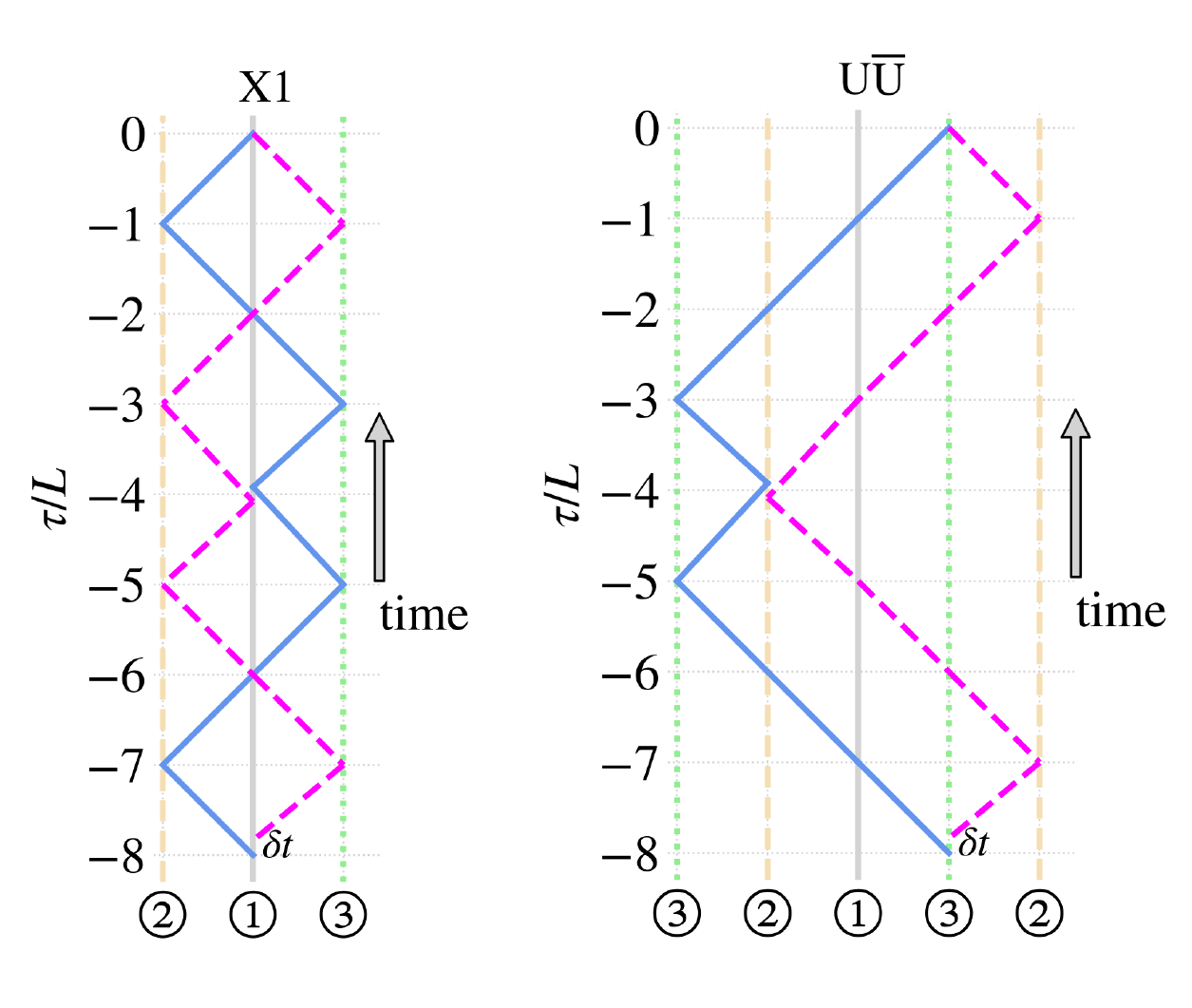}
\caption{\label{fig:2nd_tdi_diagram_8L} The geometric diagrams of X1 and $\mathrm{U\overline{U}}$ \cite{Wang:2011,Wang:2020pkk}. The vertical lines represent the spacecraft trajectories over time, with \textcircled{$i$} denoting S/C$i$ ($i = 1, 2, 3$). Additional trajectories for S/C2 and S/C3 are plotted to avoid the crossings at noninteger delay time points. The blue solid lines depict the path of one virtual laser beam, while the magenta lines represent the path of another beam. (Diagrams reused from \cite{Wang:2024alm}.) }
\end{figure}

In the second-generation TDI Michelson configuration, each observable utilizes four interferometric links from two arms. By selecting different initial S/C and sequences, three ordinary observables (X1, Y1, Z1) are defined as follows:
\begin{align} 
{\rm X1:} \ & \overrightarrow{121313121} \ \overleftarrow{131212131}, \label{eq:X1_measurement} \\
{\rm Y1:} \ & \overrightarrow{232121232} \ \overleftarrow{212323212}, \\
{\rm Z1:} \ & \overrightarrow{313232313} \ \overleftarrow{323131323}.
\end{align}
These expressions follow the notation in \cite{Vallisneri:2005ji}, where arrows represent the temporal order: "$\rightarrow$" indicates the forward-time direction from left to right, while "$\leftarrow$" denotes the backward-time direction from left to right. The numbers correspond to the indices of the S/C. For instance, the term $\overrightarrow{121313121}$ in Eq. \eqref{eq:X1_measurement} describes the path S/C1 $\rightarrow$ S/C2 $\rightarrow$ S/C1 $\rightarrow$ S/C3 $\rightarrow$ S/C1 $\rightarrow$ S/C3 $\rightarrow$ S/C1 $\rightarrow$ S/C2 $\rightarrow$ S/C1, as illustrated by the blue solid lines in the left diagram of Figure \ref{fig:2nd_tdi_diagram_8L}. The term $\overleftarrow{131212131}$ in Eq. \eqref{eq:X1_measurement} represents the path traced by the dashed magenta lines, where the sequence from left to right moves from the latest to the earliest time point. In this work, we focus on the second-generation TDI schemes. For brevity, we refer to the Michelson configuration without explicitly reiterating the second-generation.

Following the same notation, three ordinary observables of hybrid Relay ($\mathrm{U\overline{U}, \ V\overline{V}, \ W\overline{W}}$) are formulated as \cite{Wang:2011,Wang:2020pkk,Wang:2024alm}:
\begin{align}
\mathrm{U\overline{U}:} \ & \overrightarrow{312323213} \ \overleftarrow{323121323}, \\
\mathrm{V\overline{V}:} \ & \overrightarrow{123131321} \ \overleftarrow{131232131}, \\
\mathrm{W\overline{W}:} \ & \overrightarrow{231212132} \ \overleftarrow{212313212}.
\end{align}
The geometric structure of $\mathrm{U\overline{U}} $ is depicted in the right diagram of Figure \ref{fig:2nd_tdi_diagram_8L}.
As illustrated, both the Michelson and hybrid Relay configurations have a total time span of $8L$ ($L \simeq 8.33$s for LISA). During the TDI process, the maximum delay corresponds to the time interval between the earliest and the latest data points. Since the earliest inter-S/C interferometric links generate data at the receiving S/C, the maximum delay is one arm shorter than the total time span. Therefore, for both Michelson and hybrid Relay, the maximum delay is $7L$ ($\tau=-7L$).

\begin{figure}[H]
\centering
\includegraphics[width=0.45\textwidth]{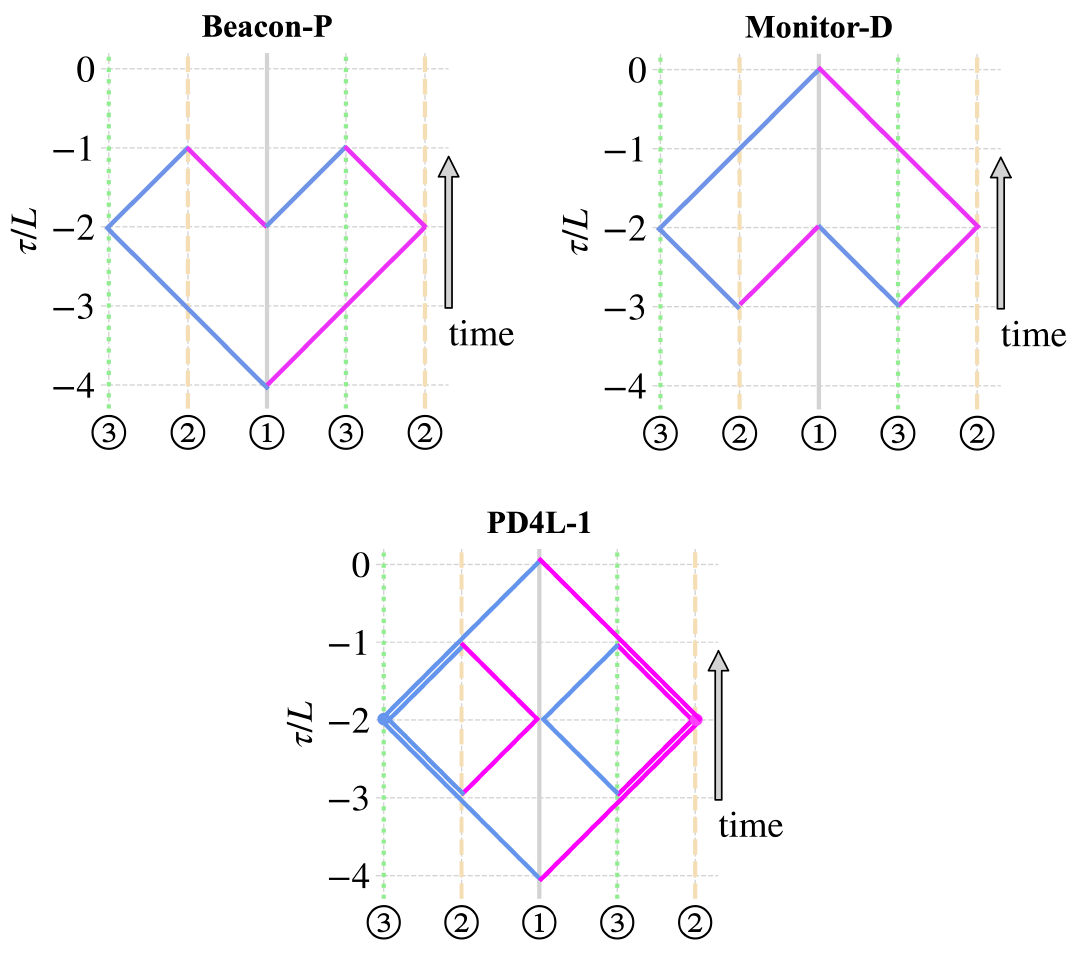}
\caption{\label{fig:tdi_diagram_4L} Geometric diagrams of the TDI channels Beacon-P and Monitor-D, and PD4L-1 \cite{Wang:2011}. The Beacon-P and Monitor-D are the first-generation TDI channels, shown in the upper panel. The lower plot illustrates PD4L-1, the first channel of the second-generation TDI PD4L configuration, which formed by combining P($t$) + D($t$). 
}
\end{figure}

In \cite{Wang:2011}, we developed a series of second-generation TDI observables by combining different first-generation TDI observables. This work focuses on identifying TDI scheme with a shorter time span. Through a systematic screening, we identify a promising scheme, denoted as PD4L (originally labeled as PE16-1bb in Eq. (7.26) of \cite{Wang:2011}), which combines the first-generation TDI Beacon (P, Q, R) and Monitor (D, E, F) configurations \cite{1999ApJ...527..814A,2000PhRvD..62d2002E}. PD4L has a shorter time span of $4L$. Another key reasons for electing PD4L is that all its observables, including its null stream, exhibit minimal null frequencies at $m/L, \ (m=1,2,3,...)$. The "4L" label distinguishes this configuration from the previously labeled PD scheme in \cite{Wang:2020pkk}, which utilizes different time shifts and time span. The Beacon-P and Monitor-D diagrams are shown in the upper panel of Figure \ref{fig:tdi_diagram_4L} \cite{Wang:2011}. The first channel of PD4L, denoted as PD4L-1, is expressed as $\mathrm{P}(t) + \mathrm{D}(t)$, with its geometric representation depicted in the lower plot of Figure \ref{fig:tdi_diagram_4L} (resembling a "Chinese knotting" pattern). The paths for three ordinary channels of PD4L are defined as follows:
\begin{align}
\textrm{PD4L-1:}\ & \overrightarrow{1 2 3 2} \ \overleftarrow{2 1 2} \ \overrightarrow{2 3 2 1} \ \overleftarrow{1 3 2 3} \ \overrightarrow{3 1 3} \ \overleftarrow{3 2 3 1}, \\
\textrm{PD4L-2:}\ & \overrightarrow{2 3 1 3} \ \overleftarrow{3 2 3} \ \overrightarrow{3 1 3 2} \ \overleftarrow{2 1 3 1} \ \overrightarrow{1 2 1} \ \overleftarrow{1 3 1 2}, \\
\textrm{PD4L-3:}\ & \overrightarrow{3 1 2 1} \ \overleftarrow{1 3 1} \ \overrightarrow{1 2 1 3} \ \overleftarrow{3 2 1 2} \ \overrightarrow{2 3 2} \ \overleftarrow{2 1 2 3}.
\end{align}

\begin{table*}[t]
\caption{\label{tab:tdi_dict}
Checklist of selected TDI channels. The first column lists the names of the TDI channels. The second column presents the paths of TDI observables, while the third column details links associated with these paths. The fourth column indicates the time span of each channel, defined as the interval from latest (virtual) interferometry time to the initial time of laser emission (where $L$ is the arm length). The fifth column specifies the maximum delay, defined as the time difference between the latest and the earliest data points required to synthesize inter-S/C interferometric measurements. The transform from ordinary channels to the optimal channels is applicable to all three TDI configurations using Eq. \eqref{eq:abc2AET}, with their corresponding optimal/orthogonal observables listed in the sixth column with subscripts. The final column displays the null frequencies in the A/E channel, where $m$ is a positive integer.
}
\begin{center}\vspace{-2mm}\small \tabcolsep 8pt
\begin{tabular*}{\textwidth}{lcccccc}
\toprule
channel & TDI path  & links & time span & max delay & optimal channels & $f_\mathrm{null}$ in A/E    \\
& & & &  & & ($m$ is positive integer) \\
\hline
X1 & $ \overrightarrow{1 2 1 3 1 3 1 2 1} \ \overleftarrow{1 3 1 2 1 2 1 3 1}$  & 16 & $8L$ & $7L$ & &  \\
Y1 & $ \overrightarrow{232121232} \ \overleftarrow{212323212} $  & 16 & $8L$ & $7L$ & (A$_\mathrm{X1}$, E$_\mathrm{X1}$, T$_\mathrm{X1}$) & $m/(4L)$  \\
Z1 & $ \overrightarrow{313232313} \ \overleftarrow{323131323}$  & 16 & $8L$ & $7L$ & &   \\
 \hline
 $\text{U}\overline{\text{U}}$ & $ \overrightarrow{31 232 3213} \ \overleftarrow{32312 1323} $ & 16 & $8L$ & $7L$ & &  \\
  $\text{V}\overline{\text{V}}$ & $ \overrightarrow{123131321} \ \overleftarrow{131232131} $ & 16 & $8L$ & $7L$ & (A$_\mathrm{U\overline{U}8L}$, E$_\mathrm{U\overline{U}8L}$, T$_\mathrm{U\overline{U}8L}$) & $m/L$   \\
 $\text{W}\overline{\text{W}}$ & $ \overrightarrow{231212132} \ \overleftarrow{212313212}$ & 16 & $8L$ & $7L$ & &  \\
\hline
 PD4L-1 & $ \overrightarrow{1 2 3 2} \ \overleftarrow{2 1 2 } \ \overrightarrow{2 3 2 1} \ \overleftarrow{1 3 2 3} \ \overrightarrow{3 1 3} \ \overleftarrow{3 2 3 1} $ & 16 & $4L$ & $3L$ & &  \\
 PD4L-2 & $ \overrightarrow{2 3 1 3} \ \overleftarrow{3 2 3} \ \overrightarrow{3 1 3 2} \ \overleftarrow{2 1 3 1} \ \overrightarrow{1 2 1} \ \overleftarrow{1 3 1 2}  $ & 16 & $4L$ & $3L$ & (A$_\mathrm{PD4L}$, E$_\mathrm{PD4L}$, T$_\mathrm{PD4L}$)  & $m/L$   \\
 PD4L-3 & $ \overrightarrow{3 1 2 1} \ \overleftarrow{1 3 1} \ \overrightarrow{1 2 1 3} \ \overleftarrow{3 2 1 2} \ \overrightarrow{2 3 2} \ \overleftarrow{2 1 2 3} $ & 16 & $4L$ & $3L$ & &   \\
 \hline
 $C^{12}_3$ & $ \overrightarrow{121} \ \overleftarrow{13} \ \overrightarrow{32} \ \overleftarrow{21} \ \overrightarrow{13} \ \overleftarrow{313} \ \overrightarrow{31} \ \overleftarrow{12} \ \overrightarrow{23} \ \overleftarrow{31} $ & 12 & $2L$ & $L$ & & $m/L$  \\
 \bottomrule
\end{tabular*}
\end{center}
\end{table*}

For the classical Sagnac and Michelson TDI configurations, three ordinary observables ($a, b, c$) could be transformed to three (quasi-)orthogonal observables (A, E, T) \cite{Prince:2002hp,Vallisneri:2007xa},
\begin{equation} \label{eq:abc2AET}
\begin{bmatrix}
\mathrm{A}  \\ \mathrm{E}  \\ \mathrm{T}
\end{bmatrix}
 =
\begin{bmatrix}
-\frac{1}{\sqrt{2}} & 0 & \frac{1}{\sqrt{2}} \\
\frac{1}{\sqrt{6}} & -\frac{2}{\sqrt{6}} & \frac{1}{\sqrt{6}} \\
\frac{1}{\sqrt{3}} & \frac{1}{\sqrt{3}} & \frac{1}{\sqrt{3}}
\end{bmatrix}
\begin{bmatrix}
a \\ b  \\ c
\end{bmatrix}.
\end{equation}
This transform is applicable when the real part of the cross-spectral density (CSD) between two ordinary channels dominates their imaginary part, as we examined in \cite{Wang:2024hgv}. Figure \ref{fig:TDI_csd_real_imag} in Appendix \ref{sec:orthogonalizatble} compares the real and imaginary components of the CSD between first two ordinary channels for PD4L, Michelson and hybrid Relay. Since the imaginary parts of PD4L's CSDs are two orders of magnitude lower than the real parts, the orthogonal transformation is expected to be valid. The A and E channels serve as science channels, effectively respond to GWs, while the T channel functions as null stream, primarily dominated by noises in the low-frequency band. To distinguish the sets of orthogonal channels derived from different TDI schemes, we append specific subscripts. Three optimal channels from the Michelson configuration are labeled as (A$_\mathrm{X1}$, E$_\mathrm{X1}$, T$_\mathrm{X1}$). To emphasis the time span, the label "$8L$" is appended to the hybrid Relay channels, (A$_\mathrm{U\overline{U}8L}$, E$_\mathrm{U\overline{U}8L}$, T$_\mathrm{U\overline{U}8L}$). The orthogonal channels from the PD4L scheme are denoted as (A$_\mathrm{PD4L}$, E$_\mathrm{PD4L}$, T$_\mathrm{PD4L}$).

In addition to the T channels formed by combining three ordinary channels, specific null streams have been developed. Notable examples include the fully symmetric Sagnac $\zeta$ in the first-generation TDI \cite{1999ApJ...527..814A} and $\zeta_1$ in the second-generation TDI \cite{Tinto:2003vj}. Additional second-generation TDI null streams are developed in \cite{Hartwig:2021mzw}. Among them, the channel $ C^{12}_3 $ exhibits minimal null frequencies and was adopted in \cite{Wang:2024hgv} to enhance instrumental noise characterization combining with science channels of hybrid Relay. Its path is expressed as follows \cite{Hartwig:2021mzw,Hartwig:2021dlc}:
\begin{equation}
C^{12}_3:  \overrightarrow{121} \ \overleftarrow{13} \ \overrightarrow{32} \ \overleftarrow{21} \ \overrightarrow{13} \ \overleftarrow{313} \ \overrightarrow{31} \ \overleftarrow{12} \ \overrightarrow{23} \ \overleftarrow{31}.
\end{equation}
The $C^{12}_3$ has a total time span of $2L$, and its geometry is illustrated in Figure 3 of \cite{Wang:2024hgv}. For clarity, a summary comparison of the three TDI schemes along with $C^{12}_3$ is provided in Table \ref{tab:tdi_dict}.

\section{TDI Noise spectra} \label{sec:noise}

Michelson TDI observables have been shown to effectively suppress laser frequency noises \cite[and references therein]{Vallisneri:2004bn,Otto:2015,Bayle:2018hnm,Wang:2020pkk,Muratore:2020mdf,Staab:2023qrb} and mitigate clock noise \cite{Otto:2012dk,Hartwig:2020tdu}. The hybrid Relay configuration has been demonstrated to achieve comparable performance in suppressing both types of noises \cite{Wang:2024alm}. The effectiveness of PD4L in laser noise suppression and clock noise mitigation is validated in Appendix \ref{sec:noise_suppression}.
In this section, we focus on the secondary noise spectra and their stability in various TDI observables, specifically considering acceleration noise and optical metrology system (OMS) noise. The noise budgets follow the specifications in \cite{2017arXiv170200786A,Colpi:2024xhw},
\begin{equation} \label{eq:noise_budgets}
\begin{aligned}
& \sqrt{ S_\mathrm{acc} } = 3 \frac{\rm fm/s^2}{\sqrt{\rm Hz}} \sqrt{1 + \left(\frac{0.4 {\rm mHz}}{f} \right)^2 }  \sqrt{1 + \left(\frac{f}{8 {\rm mHz}} \right)^4 }, \\
& \sqrt{ S_\mathrm{oms} } = 15 \frac{\rm pm}{\sqrt{\rm Hz}} \sqrt{1 + \left(\frac{2 {\rm mHz}}{f} \right)^4 }.
 \end{aligned}
\end{equation}

\begin{figure*}[htp]
\centering
\includegraphics[width=0.98\textwidth]{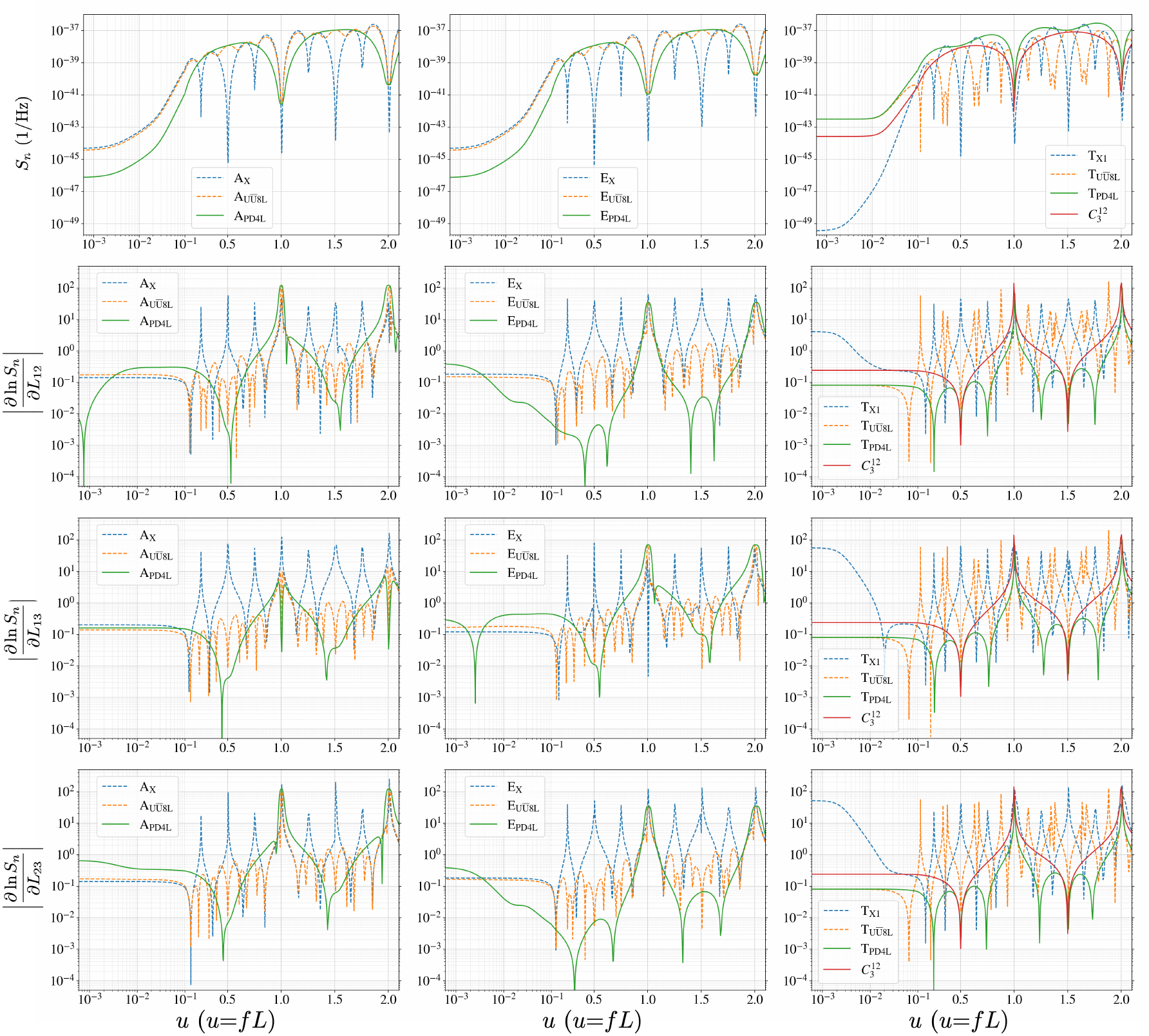}
\caption{\label{fig:Sn_dSn_dLij_AET} Noise PSDs (first row) and their derivatives with respect to three arm lengths (second to fourth rows) for the orthogonal observables from three TDI schemes. Columns correspond to the A, E, and T channels from left to right. The specific null stream $C^{12}_3$ is also shown in the third column for comparison with other T channels. As indicated in the first row, the A and E channels are identical within each TDI configuration. The Michelson observables exhibit null frequencies at $u=m/4 \ (m=1,2,3...)$, while the science channels of hybrid Relay and PD4L, as well as T$_\mathrm{PD4L}$ and $C^{12}_3$, show nulls at $u=m$. 
The second to fourth rows show the derivatives with respect to the $L_{12}$, $L_{13}$, and $L_{23}$, respectively. Despite identical PSDs, the A and E channels respond differently to changes in arm lengths. The largest variations occur near the null frequencies due to the vanishing PSD values. Consequently, the Michelson configuration (dashed blue curves) exhibits the highest instability at high frequencies. The science channels of hybrid Relay (dashed orange curves) show improved stability, benefiting from fewer null frequencies. The PD4L observables (solid green curves) display frequency- and arm-dependent stability. At low frequencies, its science channels are relatively unstable compared to the other two configurations. At high frequencies, their fluctuations could also be higher than those of hybrid Relay near null frequencies; however, A$_\mathrm{PD4L}$ and E$_\mathrm{PD4L}$ can achieve greater stability at certain frequencies. Notably, E$_\mathrm{PD4L}$ is particularly stable under changes in $L_{12}$ and $L_{23}$. 
Among the null streams (right column), T$_\mathrm{PD4L}$ is the most robust, while $C^{12}_3$ is generally stable but slightly less so. T$_\mathrm{X1}$, in contrast, is highly unstable and susceptible to performance degradation across the band.
}
\end{figure*}

Assuming equal arm lengths and identical instrumental noise contributions from all six interferometric links, the analytical noise spectral densities for Michelson TDI observables are given by \cite{Krolak:2004xp,Vallisneri:2012np}:
\end{multicols}
\begin{align}
\centering
S_\mathrm{X1} = & 256 \sin ^4 x \cos^2 x (\cos 2 x+3) S_\mathrm{acc}
                                    + 256 \sin ^4 x \cos ^2 x S_\mathrm{oms}, \\
S_\mathrm{X1-X2} = &  -64 \sin x \sin^3 2 x  S_\mathrm{acc} 
                                     -16 \sin x \sin^3 2 x S_\mathrm{oms}, \\ 
S_\mathrm{A_{X1}}  = S_\mathrm{E_{X1}} = &  256 \sin^4 x \cos^2 x (2 \cos x + \cos 2 x+3)  S_\mathrm{acc}  + 128 \sin^4 x \cos^2 x (\cos x+2) S_\mathrm{oms}, \\
S_\mathrm{T_{X1}}  = & 2048 \sin^4 \frac{x}{2}  \sin^4 x \cos^2 x S_\mathrm{acc} + 512 \sin^2 \frac{x}{2} \sin^4 x \cos^2 x S_\mathrm{oms}.
\end{align}
where $x=2 \pi f L$. The second equation corresponds to the CSD between channels X1 and X2. The corresponding noise spectra for the hybrid Relay channels are given by:
\begin{align}
\centering
S_\mathrm{U\overline{U}} = & 4 (11 -4 \cos x+\cos 2 x - 8 \cos 3 x  - 4 \cos 5 x + 3 \cos 6 x + \cos 8 x ) S_\mathrm{acc} \notag \\
              & + 4 ( 4 - \cos x - \cos 2 x  - 2 \cos 3 x - \cos 5 x + \cos 6 x)  S_\mathrm{oms}, \\
S_\mathrm{U\overline{U}-V\overline{V}} = & -256 \sin ^4 \frac{x}{2} \cos ^2 \frac{x}{2} (8 \cos x +6 \cos 2 x + 3 \cos 3 x + \cos 4 x + 5) S_\mathrm{acc} \notag  \\
                                   & + 8 \sin ^2 x ( \cos x + \cos 3 x + \cos 4 x + \cos 5 x -1) S_\mathrm{oms}, \\ 
S_\mathrm{A_{U\overline{U}8L}}  = S_\mathrm{E_{U\overline{U}8L}} = & \left( 56 -12 \cos x - 36 \cos 3x  - 12 \cos 4x  - 20 \cos 5x + 16 \cos 6x + 4 \cos 7x + 4 \cos 8x \right) S_\mathrm{acc} \notag \\
                                   & + \left( 20 - 4 \cos x - 6 \cos 2x - 8 \cos 3x - 4 \cos 4x - 6 \cos 5x + 6 \cos 6x + 2 \cos 7x \right) S_\mathrm{oms}, \\
S_\mathrm{T_{U\overline{U}8L}}  = & 128 \sin ^4 \frac{x}{2} ( \cos x + \cos 2 x + \cos 3 x )^2 S_\mathrm{acc} + 8 \left( \sin \frac{x}{2} - \sin \frac{7 x}{2} \right)^2 S_\mathrm{oms}.
\end{align}
The noise spectra of PD4L channels are:
\begin{align}
\centering
S_\mathrm{PD4L1} = & 64 \sin ^4 \frac{x}{2} (4 \cos x + \cos  2 x +5) S_\mathrm{acc}
                                    + 32 \sin ^2  \frac{x}{2}  (\cos x +2)  S_\mathrm{oms}, \\
S_\mathrm{PD4L1-PD4L2} = & 16 \sin ^2 \frac{x}{2}  (4 \cos x +5) S_\mathrm{acc} 
                                    + 8 \sin ^2 x (\cos x +2) S_\mathrm{oms}, \\ 
S_\mathrm{A_{PD4L}}  = S_\mathrm{E_{PD4L}} = & 64 \sin ^4 \frac{x}{2}  (2 \cos x+\cos 2 x+3) S_\mathrm{acc} +32 \sin ^4 \frac{x}{2} (\cos x +2) S_\mathrm{oms}, \\
S_\mathrm{T_{PD4L}}  = & 128 \sin ^4 \frac{x}{2}  (\cos x+2)^2 S_\mathrm{acc}
                                    +8 \left(3 \sin \frac{x}{2}+\sin \frac{3x}{2} \right)^2 S_\mathrm{oms}.
\end{align}
\begin{multicols}{2}
The noise PSDs of the optimal channels from three TDI schemes are shown in the top row of Figure \ref{fig:Sn_dSn_dLij_AET}. The plots are ordered left to right by the A, E, and T channels. The calculations are performed under a static unequal-arm setup token from a random time point along a numerical orbit \footnote{https://github.com/gw4gw/LISA-Like-Orbit} \cite{Wang:2017aqq}, with instantaneous arm length (in seconds): $[ L_{12}, L_{21}, L_{13}, L_{31}, L_{23}, L_{32}]  \simeq [8.296, 8.297, 8.275, 8.277, 8.313, 8.313]$ s.
To clearly highlight the locations of null frequencies, the x-axis is defined as $u = fL$, a dimensionless frequency variable (where $f$ is frequency and $L$ is the nominal arm length). A logarithmic scale is applied for $u < 0.1$, while a linear scale is used for higher values. As seen in the first two plots, the science channels of Michelson configuration exhibit nulls at $u=m/4$, where $m$ is a positive integer. In contrast, the science channels, A$_\mathrm{U\overline{U}8L}$/E$_\mathrm{U\overline{U}8L}$ and A$_\mathrm{PD4L}$/E$_\mathrm{PD4L}$, have nulls only at $u=m$.
The top-right plot shows the PSDs of the null streams from three TDI schemes, as well as $C^{12}_3$. Among these, T$_\mathrm{PD4L}$ and $C^{12}_3$ exhibit the fewest nulls, occurring only at $u=m$. The T channel of Michelson (T$_\mathrm{X1}$) shows nulls at $u=m/4$, while T$_\mathrm{U\overline{U}8L}$ exhibits more frequent null frequencies.

The noise spectra of TDI observables depend on both the instrumental noise components and arm lengths. Although the instrumental noises are assumed to be stationary, the noise PSDs vary over time due to changes in arm lengths driven by orbital dynamics. 
To assess the stabilities of noise spectra, we compute their logarithmic derivatives with respect to the three arm lengths. The second to fourth rows in Figure \ref{fig:Sn_dSn_dLij_AET} present the derivatives with respect to $L_{12}$, $L_{13}$, and $L_{23}$, respectively. In these calculation, we assume the deviations in $L_{ij}$ and $L_{ji}$ are equal. Thus, $\frac{\partial \ln S_n}{\partial L_{ij}}$ incorporates the effects from both links, although they could be slightly different in practice.
The derivatives generally exhibit sharp peaks around their null frequencies, where the PSDs are particularly small and highly sensitive to changes in arm lengths. The downward spikes in some curves primarily result from sign reversals when taking the absolute values. 

The A and E channels are shown in the first two columns of Figure \ref{fig:Sn_dSn_dLij_AET}. In the low-frequency regime ($u < 0.1$), the observables from Michelson and hybrid Relay configurations exhibit more stable behavior than those of PD4L. Their derivatives remain below 0.2 for all three arms (dashed curves). In contrast, the PD4L shows greater sensitivity to arm length changes in this frequency range. This is partly because its shorter time span leads to lower noise PSDs in low-frequency band (as seen in the first row of Figure \ref{fig:Sn_dSn_dLij_AET}), making the relative deviations more significant.
In the high frequency regime ($u>0.1$), the stability of the TDI channel becomes more configuration-dependent. The Michelson observables are less favored due to their susceptibility to the numerous null frequencies at $u=m/4$. The science channels of hybrid Relay are more stable overall, with derivatives magnitudes typically ranging from 0.1 to 1, except near the null frequencies. The stabilities of PD4L channels vary with both frequencies and arms. Although fluctuations around the null frequencies can be more pronounced than those of the hybrid Relay, the PD4L channels may exhibit better stability than hybrid Relay at certain frequencies between null frequencies. Notably, E$_\mathrm{PD4L}$ remains highly stable when the variations occurs in $L_{12}$ and $L_{23}$.

The behaviors of the null streams are illustrated in the third column of Figure \ref{fig:Sn_dSn_dLij_AET}. Among these, T$_\mathrm{X1}$ (dashed blue curves) is significantly unstable with large derivative magnitudes, especially at low frequencies.
Such instability may compromise data analysis accuracy as investigated in \cite{Wang:2020fwa,Wang:2024alm,Wang:2024hgv}. 
Among other null streams, T$_\mathrm{PD4L}$ (solid green) is the most stable across the target frequency range (except at null frequencies), while T$_\mathrm{U\overline{U}8L}$ (dashed orange) is stable for $u< 0.1$ but degrades at higher frequencies. The null stream $C^{12}_3$ (solid red) exhibits overall stability across the spectrum but is slightly less robust than the T$_\mathrm{PD4L}$.

\section{GW Response in TDI Observables} \label{sec:gw_simulation}

\begin{figure*}[htbp]
\centering
\includegraphics[width=0.9\textwidth]{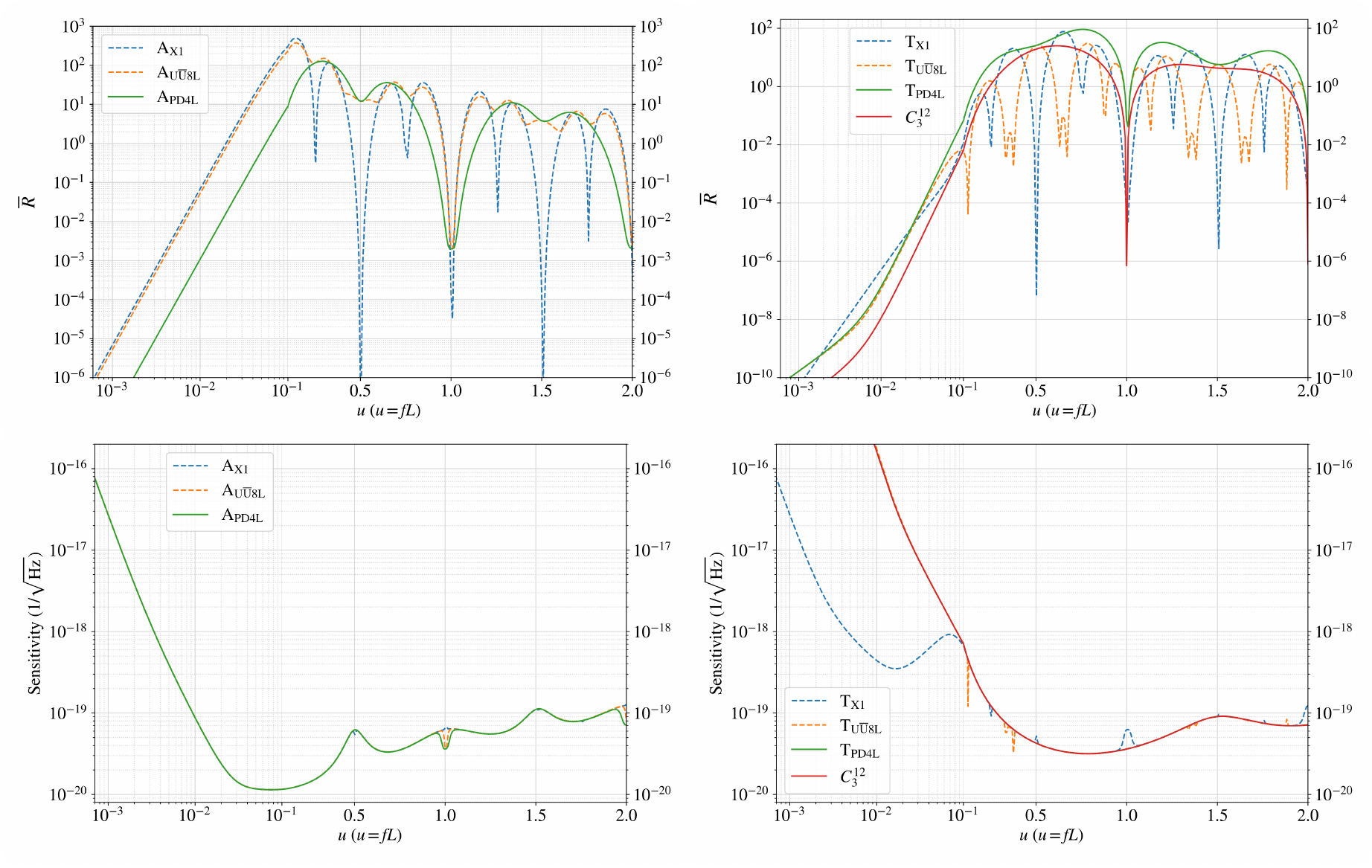}
\caption{\label{fig:resp_sensitivity} Sky-averaged GW responses (upper) and sensitivities (lower) of selected TDI channels. The left column presents results for the science A channels, while the right column depicts the null streams. The spikes in the sensitivity curves are caused by numerical error near the null frequencies. 
In the bottom-right plot, the curves for T$_\mathrm{U\overline{U}8L}$, T$_\mathrm{PD4L}$ and $C^{12}_3$ largely overlap.
}
\end{figure*}

The sky-averaged GW response of a TDI channel is computed using:
\begin{equation} \label{eq:resp_averaged}
\begin{aligned}
 \mathcal{R}_{\rm TDI} (f) =& \frac{1}{4 \pi}  \int^{2 \pi}_{0} \int^{\frac{\pi}{2}}_{-\frac{\pi}{2}} |F^{\rm{GW}}_{ \rm TDI} (f)|^2 \cos \beta {\rm d} \beta {\rm d} \lambda,
\end{aligned}
\end{equation}
where $(\lambda, \beta)$ represent the ecliptic longitude and latitude. To enable better comparison across frequencies, the responses are normalized by $u^2$, defined as $\overline{R} = \mathcal{R}_{\rm TDI} / u^2$, and are shown in the upper panel of Figure \ref{fig:resp_sensitivity}. Corresponding average sensitivities are calculated as $ \sqrt{ \frac{S_{n, \mathrm{TDI}}}{ \mathcal{R}_\mathrm{TDI} } } $ and displayed in the lower panel of the figure.
The left and right columns show the science A channels and the null channels, respectively. Since the A and E channels have identical average responses and sensitivities, only A channels are shown for brevity (the response/sensitivity of A and E could be slightly different under unequal-arm case, we ignore the difference here).
As illustrated in the upper left plot, the average responses of A$_\mathrm{U\overline{U}8L}$ and A$_\mathrm{PD4L}$ are smoother than the Michelson at higher frequencies ($u > 0.1$), owing to fewer null frequencies. At lower frequencies ($u<0.1$), the response of A$_\mathrm{PD4L}$ is roughly two orders of magnitude lower than that of A$_\mathrm{X1}$ or A$_\mathrm{U\overline{U}8L}$. This implies stronger GW signal suppression in the PD4L channels, primarily due to the stacking of links over a shorter time span, which leads to increased cancellation of the signal in the long-wavelength regime.
Interestingly, as shown in the lower left plot, the average sensitivities of all science channels are essentially identical. The reduced GW response in A$_\mathrm{PD4L}$ channel is compensated by its lower noise PSD (shown in the first plot of Figure \ref{fig:Sn_dSn_dLij_AET}), resulting in an equal average sensitivity. However, implementing PD4L scheme requires higher-precision interpolation than the Michelson or hybrid Relay. The current algorithm should meet the requirement \cite{Shaddock:2004ua,Staab:2024nyo}.

The right column of Figure \ref{fig:resp_sensitivity} shows the GW responses and sensitivities of four null streams. As seen in the upper plot, the T$_\mathrm{PD4L}$ (solid green) and $C^{12}_3$ (solid red) exhibit the smoothest responses across the frequency band, with T$_\mathrm{PD4L}$ consistently showing higher amplitude the $C^{12}_3$. In contrast, T$_\mathrm{X1}$ (dashed blue) and T$_\mathrm{U\overline{U}8L}$ (dashed orange) show prominent fluctuations at high frequencies due to their dense null frequency structure.
The lower right plot displays the corresponding sensitivities. T$_\mathrm{X1}$ appears more sensitive at low frequencies for $u < 0.1$. However, this improvement is correlated with the E$_\mathrm{X1}$ channel as analyzed in \cite{Hartwig:2023pft}. The other three null channels exhibit nearly identical sensitivities, with their curves closely overlapping. 

\begin{figure*}[htp]
\centering
\includegraphics[width=0.85\textwidth]{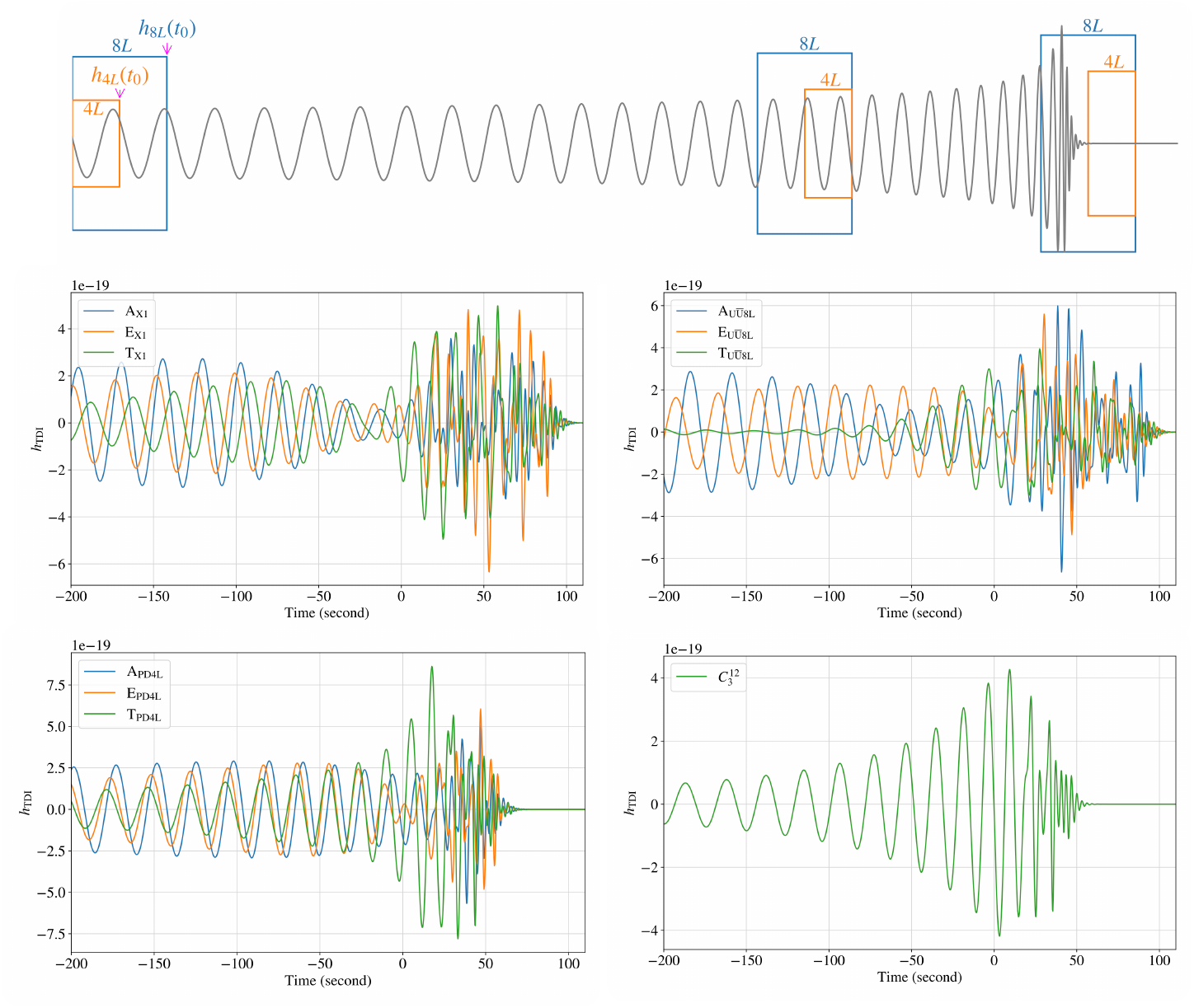}
\caption{\label{fig:TDI_waveform_td} A merging GW signal from a MBBH ($m_1=3 \times 10^4 M_\odot$, $m_2=1 \times 10^4 M_\odot$ at $z=0.2$) and the corresponding waveforms in different TDI channels. The top row shows the original waveform. Blue and orange frames indicate the TDI operation time span for $8L$ (Michelson and hybrid Relay) and $4L$ (PD4L) configurations, respectively. From left to right, three frame pairs illustrate: 1) margin effects at data boundaries, 2) frequency aliasing at high frequencies, and 3) tails effects at the signal's end. (Note: realistic TDI involves GW responses across multiple inter-S/C links.) The middle and bottom rows display the responded waveforms in selected TDI channels. The merger time arriving at the solar-system barycenter is set to be $t=0$, and the signal arrives later at the detector in current simulation setup. The middle left plot depicts the waveforms in the Michelson optimal channels (A$_\mathrm{X1}$, E$_\mathrm{X1}$, T$_\mathrm{X1}$), the middle right panel displays those from hybrid Relay (A$_\mathrm{U\overline{U}8L}$, E$_\mathrm{U\overline{U}8L}$, T$_\mathrm{U\overline{U}8L}$). The bottom left plot illustrates the waveforms in PD4L observables, and the bottom right shows the null stream $C^{12}_3$. These waveforms demonstrate that shorter-span TDI observables produce smoother waveforms with reduced signal tails.
}
\end{figure*}

The sky-averaged responses and sensitivities are derived under assumption of monochromatic signals at each frequency. To assess how TDI observables respond to evolving signal, we apply TDI operations to a rapidly chirping waveform from a MBBH system with component masses $m_1=3 \times 10^4 M_\odot$ and $m_2=1 \times 10^4 M_\odot$ at redshift $z=0.2$. The goal is to evaluate the performance of various TDI observables, particularly in the high frequency regime. The top row of Figure \ref{fig:TDI_waveform_td} displays the original merging waveform. Blue and orange frames indicate the time span of TDI operation for $8L$ (used in Michelson or hybrid Relay) or $4L$ (used in PD4L), respectively. The three pairs of frames, from left to right, highlight the following key effects:
\begin{itemize}
\item Margin Effect: TDI channels with shorter span begin generating data earlier, thereby reducing margin loss at data boundaries. For instance, Michelson TDI generates its first data point after $\sim 8L$, while PD4L advances this to $\sim 4L$. This feature is particularly advantageous for handling observational interruptions, as discussed in \cite{Seoane:2021kkk,Castelli:2024sdb}, , and becomes even more beneficial in the presence of frequent data gaps.
\item Frequency Aliasing, longer-span TDI operations traverse more of the waveform's evolution, making them more susceptible to frequency aliasing, especially in the high-frequency region.
\item Tail Effects, shorter-span TDI observables pass through the end of waveform more quickly, resulting in shorter signal tails. 
\end{itemize}
Note: actual TDI operation involves GW responses accumulated across multiple inter-S/C links, not just the idealized waveform shown here.

The resulting waveforms in the orthogonal channels of Michelson, hybrid Relay and PD4L, as well as the null stream $C^{12}_3$, are shown in the middle and bottom rows of Figure \ref{fig:TDI_waveform_td}. The merge time arriving at solar-system barycenter is set to $t=0$. Due to the specific source and detector geometry, the signal arrives later at the detector.
Comparing the waveforms in these TDI channels with different time spans, the waveforms of Michelson observables (middle left plot) exhibit irregular phases and amplitudes variations near merging, primarily due to aliasing and modulations around null frequencies. The waveform of hybrid Relay (middle right plot) are less affected by amplitude modulation compared to Michelson, benefiting from fewer null frequencies. However, the phases remain arrhythmic due to mixed phase contributions across different periods. 
In contrast, the waveforms in PD4L observables (bottom left plot) are noticeably smoother, because the shorter TDI span mitigates aliasing effects and produce cleaner phase evolution with shorter tails. With minimal null frequencies in all three orthogonal channels, the PD4L waveforms maintain better phase coherence.
In the bottom right plot, the null stream $C^{12}_3$ exhibits the smoothest phase progression and the shortest tail, benefiting from its minimal null frequencies and compact $2L$ time span.

These comparisons highlight that shorter-span TDI configurations are better suited for handling high-frequency signals. They produce smoother time-domain waveforms, which facilitate signal modeling in the frequency-domain. In the following section, we perform parameter inference using this chirping signal and compare the performance across different TDI configurations.

\section{Parameter inference with simulated data} \label{sec:parameters_inference}

In this section, we perform two types of parameter inferences: 1) signal parameter estimation: to evaluate the ability of TDI configurations to analyze high-frequency and low-frequency GW signals; and 2) noise characterization: to estimate the amplitudes of noises components using pure noise data of PD4L channels. The inferences are conducted using simulated data with hybrid Relay and/or PD4L configurations. Time-domain waveform and noise data for each ordinary TDI observable are generated using \textsf{SATDI} \cite{Wang:2024ssp}. The noise incorporates acceleration noise and OMS noise, both assumed to be Gaussian and stationary. For each TDI configuration with three ordinary data streams, optimal data streams are obtained by applying the (quasi-)orthogonal transformation defined in Eq. \eqref{eq:abc2AET}.

To infer binary parameters, we simulate two GW signals from spinless MBBH. The first binary, with component masses $m_1 = 3 \times 10^4 M_\odot$ and $m_2 = 10^4 M_\odot$ in the source frame, produces a waveform in the high-frequency band, as illustrated in Figure \ref{fig:TDI_waveform_td}. The second binary, with $m_1=6 \times 10^5 \mathrm{M}_\odot$ and $m_2=2 \times 10^5 \mathrm{M}_\odot$ at the same location, generates a lower frequency signal. Both sources are placed at a redshift of $z=0.2$ which corresponds to a luminosity distance of $d_L =  1012.3$ Mpc \cite{Planck:2018vyg,Astropy:2013muo}. The distance is chosen to be closer than realistic astronomical expectations, in order to amplify the signal, increase inference precision, and better reveal the performance differences between TDI configurations.
The source is located at ecliptic longitude $\lambda=4.6032$ rad and ecliptic longitude $\beta=\pi/10$ rad. The polarization angle is set to $\psi=0.55$ rad, and the inclination angle between the orbital axis and the line-of-sight is $\iota=\pi/3$. The merger time at  the solar-system barycenter is defined as $t_c = 0$, with a reference phase $\phi_c$. The time-domain waveform is generated using the SEOBNRv4HM model \cite{Bohe:2016gbl}. During inference, we estimate the redshifted chirp mass $\mathcal{M}_c = (m_1 m_2)^{3/5}/(m_1 + m_2)^{1/5}$ and the mass ratio $q=m_2/m_1$, rather than the individual component masses. We also infer the merger time offset $\Delta t_c$ relative to $t_c$.
The frequency-domain waveform is modeled using a reduced-order implementation of SEOBNRv4HM \cite{Cotesta:2020qhw}. A total nine parameters are inferred: [$\mathcal{M}_c$, $q$, $d_L$, $\iota$, $\lambda$, $\beta$, $\psi$, $\Delta t_c$, $\phi_c$].

To characterize noise, we infer 12 parameters that describe the amplitudes of acceleration noises and OMS noises. As specified in Eqs. \eqref{eq:noise_budgets}, the amplitudes for the six acceleration noise components are set to $A_{\mathrm{acc}ij} = 3$, where $ij$ represents the spacecraft pair with S/C$i$ facing S/C$j$. For OMS noise, the six parameters for amplitudes are $A_{\mathrm{oms}ij} = 15$. In the inference, their squared amplitudes are treated as free parameters. Priors for the squared acceleration noise amplitude, $A^2_\mathrm{acc}$, are uniformly distributed within the range of [0, 40], and those for the OMS noises, $A^2_\mathrm{oms}$, are chosen in the range [50, 300]. The inference is carried out in the frequency-domain, using a low-frequency cutoff of 0.03 mHz and a high-frequency cutoff of 0.1 Hz. 

Parameter estimation is performed using the Bayesian framework. The likelihood function is defined as \cite{Adams:2010vc,Romano:2016dpx}:
\begin{equation} \label{eq:likelihood_fn}
\begin{split}
\ln \mathcal{L} (d|\vec{\theta} ) = \sum_{f_i} & \left[  
-\frac{1}{2} \tilde{\mathbf{n}}^T \mathbf{C}^{-1}  \tilde{\mathbf{n}}^\ast   - \ln \left( \det{ 2 \pi \mathbf{C} } \right)
   \right].
\end{split}
\end{equation}
where $\tilde{\mathbf{n}}$ is the frequency-domain noise data vector, and $\mathbf{C}$ is the noises covariance matrix for the three orthogonal channels:
\begin{equation} \label{eq:cov_mat}
\begin{aligned}
 \mathbf{C} = & \frac{T_\mathrm{obs}}{4}
 \begin{bmatrix}
S_\mathrm{AA} & S_\mathrm{AE} & S_\mathrm{AT} \\
S_\mathrm{EA} & S_\mathrm{EE} & S_\mathrm{ET} \\
S_\mathrm{TA} & S_\mathrm{TE} & S_\mathrm{TT}
\end{bmatrix},
\end{aligned}
\end{equation}
where $T_\mathrm{obs}$ is the data duration. For MBBH analysis, a modified formulation with matched filtering is employed to remove the impact of random noise fluctuations. Sampling is conducted using the nested sampling algorithm implemented in \textsf{MultiNest} \cite{Feroz:2008xx,Buchner:2014nha}. 

\begin{figure*}[htbp]
\centering
\includegraphics[width=0.95\textwidth]{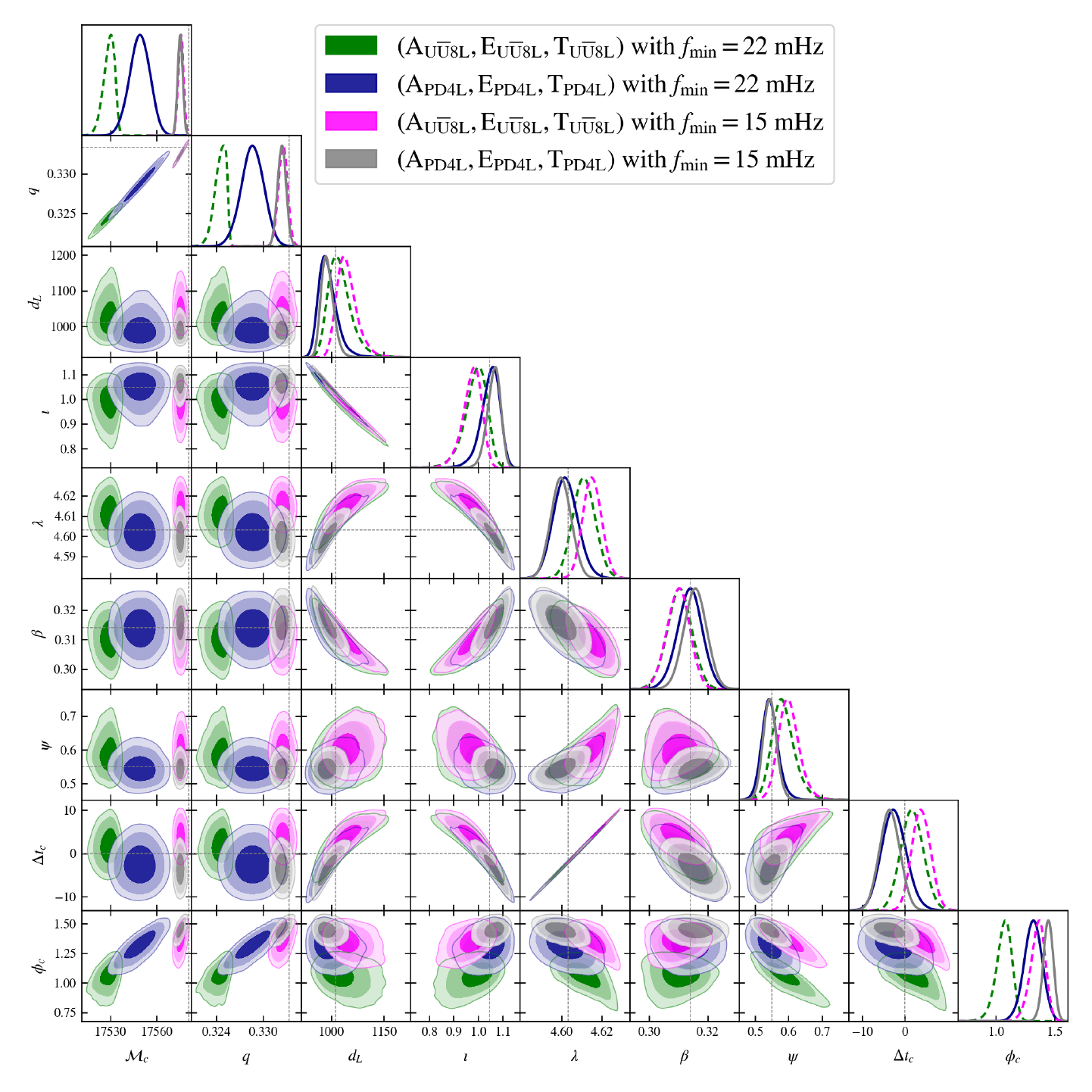}
\caption{\label{fig:corner_mbbh} Posterior parameter distributions inferred using the hybrid Relay (A$_\mathrm{U\overline{U}8L}$, E$_\mathrm{U\overline{U}8L}$, T$_\mathrm{U\overline{U}8L}$) and PD4L (A$_\mathrm{PD4L}$, E$_\mathrm{PD4L}$, T$_\mathrm{PD4L}$) TDI schemes for a GW signal from a coalescing MBBH with $(m_1=3 \times 10^4, m_2 = 10^4) \mathrm{M}_\odot$ at redshift $z=0.2$. The green and blue contours show results obtained with a low-frequency cutoff of $f_\mathrm{min}=22$ mHz using hybrid Relay and PD4L, respectively. The magenta and grey colors indicate the results with a lower cutoff of $f_\mathrm{min}=15$ mHz for these two TDI schemes. A high frequency cutoff of 0.4 Hz is applied, covering the post-merger phase. The signal spans for $\sim$1700 seconds (from 22 mHz to 0.4 Hz) and extends to 5000 seconds when starting from 15 mHz, both in the SSB frame. Shaded gradients represent the $1\sigma$, $2\sigma$, and $3\sigma$ confidence regions, with darker shades indicating higher confidence. Grey dashed lines indicate the true (injected) parameter values. To better reveal the limitations of TDI configurations at high frequencies, the noise PSD is artificially reduced by a factor of 10 in this inference.
}
\end{figure*}

The hybrid Relay (A$_\mathrm{U\overline{U}8L}$, E$_\mathrm{U\overline{U}8L}$, T$_\mathrm{U\overline{U}8L}$) and PD4L (A$_\mathrm{PD4L}$, E$_\mathrm{PD4L}$, T$_\mathrm{PD4L}$) configurations are employed to estimate the parameters of MBBH sources. Figure \ref{fig:corner_mbbh} presents the inferred posterior distributions for high frequencies GW signal from the lighter MBBH. Two frequency ranges are implemented for each TDI scheme: one with a frequency span from 22 mHz to 0.4 Hz (yielding a $\sim$1700 seconds signal duration in the SSB frame), and the other using a lower frequency cutoff of 15 mHz, extending the signal duration to 5000 seconds. To sufficiently reveal the deficiency of TDI scheme at high frequencies, the noise PSDs is artificially reduced by a factor of 10 in this test.

Distributions obtained from the shorter duration data are shown in green (hybrid Relay) and blue (PD4L). Two configurations result in noticeably different distributions across both intrinsic and extrinsic parameters. Neither configuration accurately recovers the intrinsic mass parameters, likely due to aliasing effects and inaccuracies in frequency-domain modeling. Nevertheless, PD4L yields distributions closer to the injected values, which we attribute to its superior mitigation of aliasing. For extrinsic parameters, such as luminosity distance, inclination, and polarization angle, PD4L again outperforms the hybrid Relay, providing more precise constraints. Additional comparisons using a mixed combination (A$_\mathrm{U\overline{U}8L}$, E$_\mathrm{U\overline{U}8L}$, $C^{12}_3$) are provided in Appendix \ref{sec:hybrid_relay_zeta}.

When the frequency cutoff is lowered to 15 mHz, the resulting distributions for the mass parameters become consistent between hybrid Relay (magenta) and PD4L (grey), as shown in in Figure \ref{fig:corner_mbbh}. This improvement is attributed to the inclusion of more inspiral phase information, which dominates the signal at lower frequencies. However, even with the extended duration, the true mass values barely included in the credible intervals, possibly due to limited faithfulness between two waveform approximants used in injection and recovery. Compared to the short-duration analysis, the longer duration leads to noticeably tighter parameter constraints as expected. While a more comprehensive evaluation across large population of source would be ideal, such analysis is computationally expensive. Based on this representative case, we conclude that PD4L outperforms the hybrid Relay in high-frequency parameter inference, supported by both waveforms diagnosis and inference results.

\begin{figure*}[htp]
\centering
\includegraphics[width=0.8\textwidth]{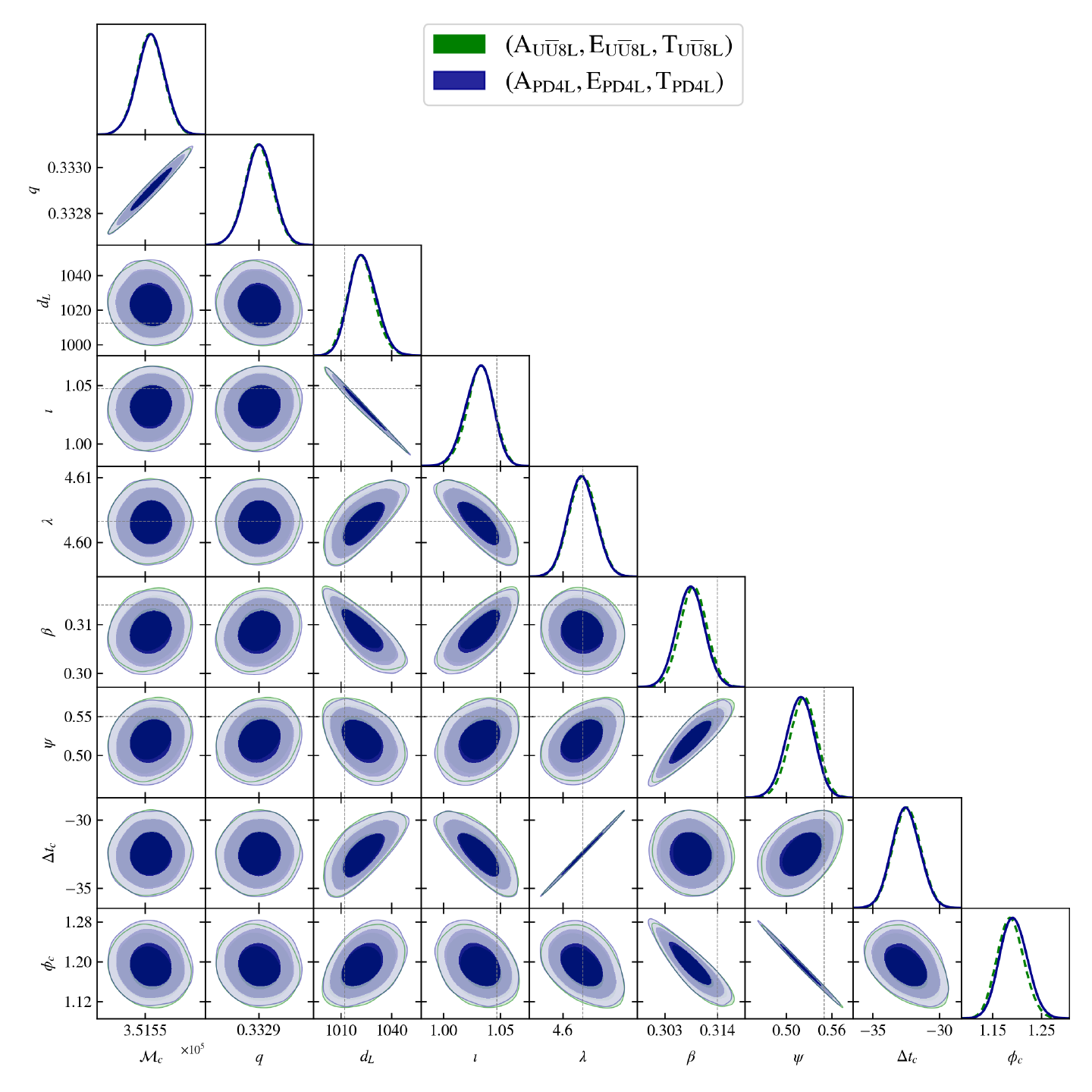}
\caption{\label{fig:corner_mbbh_heavy} 
Posterior parameter distributions inferred using the hybrid Relay (A$_\mathrm{U\overline{U}8L}$, E$_\mathrm{U\overline{U}8L}, $T$_\mathrm{U\overline{U}8L}$) and PD4L (A$_\mathrm{PD4L}$, E$_\mathrm{PD4L}$, T$_\mathrm{PD4L}$) schemes for a GW signal from a coalescing MBBH with component masses $(m_1=6 \times 10^5, m_2=2 \times 10^5) \mathrm{M}_\odot$ at redshift $z=0.2$. The analysis utilizes the frequency band [1.1, 28] mHz, during which the signal evolves over 9.5 hours in the SSB frame. The green contours represent the posterior distribution from the hybrid Relay scheme, while the blue contours correspond to those from the PD4L. The near-complete overlap between the two sets of distributions indicates that both TDI configurations perform equivalently for massive binaries in the low frequency regime.
}
\end{figure*}

The inferred distribution with hybrid Relay and PD4L configurations for the heaver MBBH system are shown in Figure \ref{fig:corner_mbbh_heavy}. The GW signal spans a frequency range from 1.1 mHz to 28 mHz and lasts 9.5 hours in the SSB frame. The results from both TDI schemes are nearly identical, with the posterior distributions fully overlapping. This suggests that both the hybrid Relay and PD4L configuration offer equivalent performance in analyzing signals in the low-frequency band, assuming adequate interpolation precision. The intrinsic mass parameters are not correctly recovered, likely also due to an insufficient match between two waveform approximants. Most extrinsic parameters, except for merger time offset $\Delta t_c$ and  reference phase $\phi_c$, successfully include the injected values within their $2\sigma$ credible intervals.

\begin{figure*}[htp]
\centering
\includegraphics[width=0.88\textwidth]{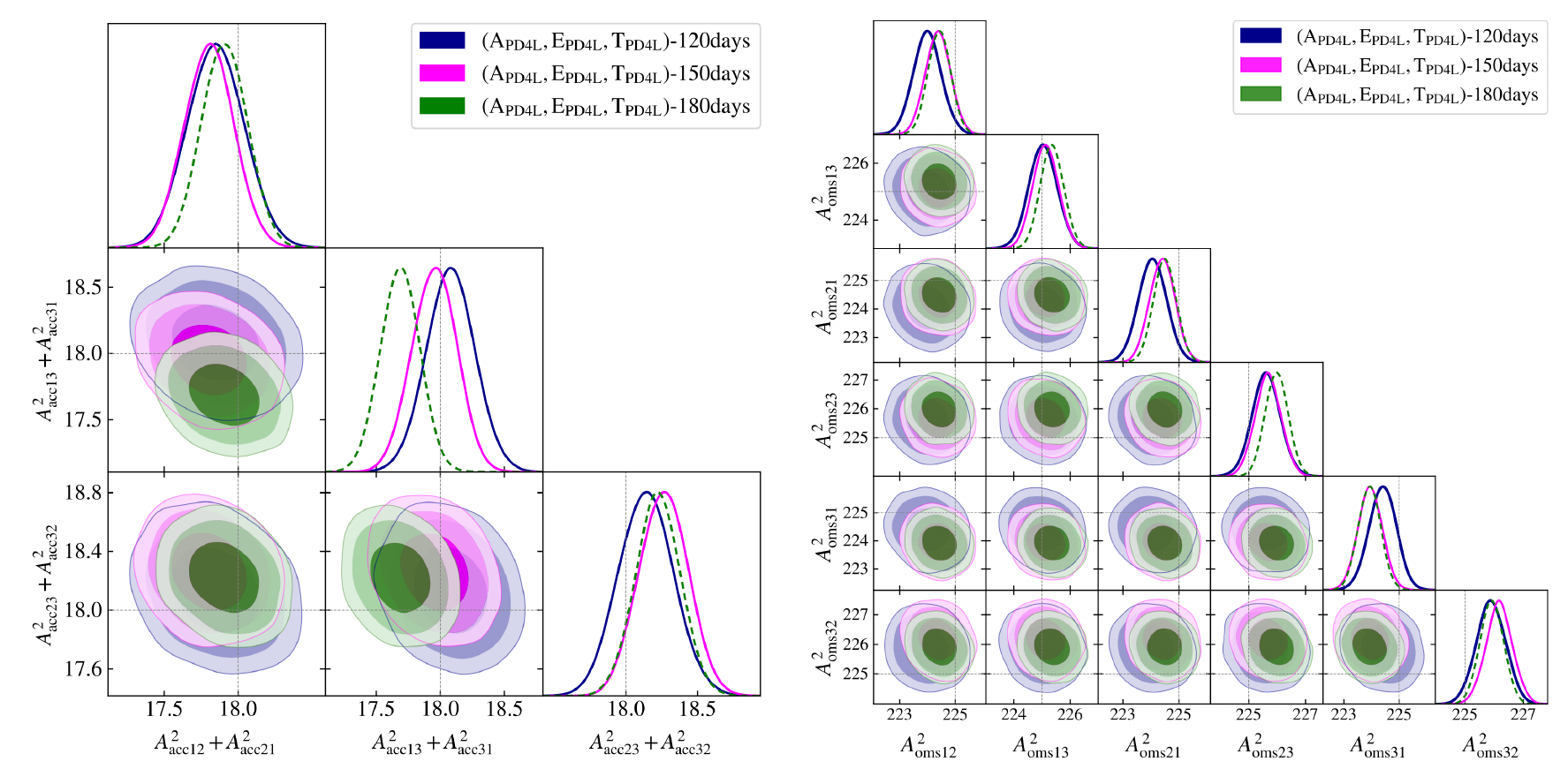}
\caption{\label{fig:corner_noises_PD} 
Inferred parameter distributions by using 120-day, 150-day, and 180-day data streams from the PD4L observables (A$_\mathrm{PD4L}$, E$_\mathrm{PD4L}$, T$_\mathrm{PD4L}$), based on a dynamically evolving unequal-arm constellation. The analysis covers a frequency range of [0.03 mHz, 0.1 Hz]. Shaded contours represent the $1\sigma$, $2\sigma$, and $3\sigma$ confidence regions, with darker shades indicating higher confidence. The grey dashed lines indicate the simulated (true) parameter values. The left plot shows the quadratic sums of acceleration noise amplitudes, and the right panel presents the squared amplitudes of OMS noise. 
}
\end{figure*}

The inferred amplitudes of noise spectra are shown in Figure \ref{fig:corner_noises_PD}. The optimal channels of PD4L (A$_\mathrm{PD4L}$, E$_\mathrm{PD4L}$, T$_\mathrm{PD4L}$) are utilized to perform the inferences over three different data durations. Results obtained from 120-day, 150-day and 180-day datasets are shown in blue, magenta, and green, respectively. Because the acceleration noises from paired test masses on an interferometric link are correlated, we present their quadratic sums, $A_{\mathrm{acc}ij} + A_{\mathrm{acc}ji} $, in the left plot. As expected, the uncertainties decrease with longer data duration. For all three durations, the values of acceleration noise parameters are effectively recovered within the $2 \sigma$ regions.
The right panel shows the distributions for the six squared OMS noise amplitudes, $A^2_{\mathrm{oms}ij}$. For the 120-day dataset, all simulated values fall within the $3 \sigma$ confidence regions. However, for the 150-day and 180-day cases, some parameters fall outside the intervals, suggesting a slight degradation in inference performance. This indicates that the noise characterization ability of the PD4L scheme is somewhat weaker than the (A$_\mathrm{U\overline{U}8L}$, E$_\mathrm{U\overline{U}8L}$, $C^{12}_3$) combination, which achieved more accurate recovery using a 180-day dataset \cite{Wang:2024hgv}. This reduced performance likely stems from the fact that the noise PSDs of A$_\mathrm{PD4L}$ and E$_\mathrm{PD4L}$ are less robust than the A$_\mathrm{U\overline{U}8L}$ and E$_\mathrm{U\overline{U}8L}$, as illustrated in Figure \ref{fig:Sn_dSn_dLij_AET}.

It is worth noting that the current simulation does not account for potential data gaps. The LISA mission is expected to operate with a duty cycle exceeding 82\% \cite{Colpi:2024xhw}. In the event that data segments are limited to durations shorter than $\sim$4 months, the PD4L scheme should still satisfy the requirement for accurate noise characterization. Giving its superior performance in analyzing GW signals at higher frequencies, PD4L remains a strong candidate for alternative TDI configuration.

\section{Conclusions and discussion} \label{sec:conclusions}

In this work, we introduce a second-generation TDI scheme, PD4L, which features minimal null frequencies and a compact operational time span of $4L$  -- half that of the fiducial Michelson or hybrid Relay with $8L$. This more compact design offers three advantages: 1) it reduces the data margin required for TDI operation; 2) it mitigates aliasing in the high frequency band; and 3) it shortens the tail at the signal end in the time domain. These advantages make the PD4L particularly well-suited for analyzing high-frequency GW signals. However, we also identify two limitations: 1) both GW signal and noise suppression are more significant in the low frequency band, which increase the need for higher interpolation accuracy when reconstructing data values; 2) the noise spectra in two science channels of PD4L are less stable than those of hybrid Relay, limiting the reliability of long-duration noise characterization.

To assess PD4L's performance, we simulated chirping GW signals and compared the response across different TDI observables. Subsequent parameter inferences results demonstrate that PD4L outperforms the hybrid Relay in the high frequency regime and performs comparably in the low frequency band. Given that the hybrid Relay has already demonstrated superiority over Michelson configuration \cite{Wang:2024alm,Wang:2024hgv}, we conclude the PD4L offers the best capability for analyzing high-frequency signals among the three TDI schemes, specifically in the band $u=fL > 0.1$. Despite the relatively lower noise stability of PD4L's science channels (than that of hybrid Relay) and its slightly limited capability for noise characterization, its null stream, T$_\mathrm{PD4L}$, exhibits superior stability compared to the $C^{12}_3$ channel, making it the most stable null stream among all configuration considered in this study. On the other hand, low-mass binaries targeted for multi-band observations with space- and ground-based detectors will remain in the higher-frequency band for a much longer duration. Their potentially low SNR may require delicate algorithms to extract the encoded information, as discussed in \cite{Wu:2025zhc}. Their GW signals are less affected by the nulls in PD4L, which will facilitate modeling compared to the Michelson configuration.

Our current GW inferences are performed in the frequency domain. However, we acknowledge that the frequency-domain model may not be optimally match the simulated signal though the Fourier transform of a finite-duration time-domain data. Tail effects and aliasing introduced by TDI operations may not be fully captured, potentially biasing parameter estimation. While more compact TDI configuration like PD4L mitigate these issues, they do not entirely eliminate them. A recent work \cite{Garcia-Quiros:2025usi} demonstrated GPU-accelerated parameter inference utilizing time-domain responses, which may offer a route to overcome the shortages in the frequency-domain models. 

In \cite{Wang:2011}, we developed a variety of TDI schemes by combining the first-generation observables, identifying some of those with minimized null frequencies across time spans of $4L$, $6L$, and $8L$. For this study, we selected the hybrid Relay and PD4L as representative configurations for the $8L$ and $4L$ spans, respectively. We present a systematic analyze of the differences and correlations among these TDI schemes in \cite{Wang:2025voa}.

\Acknowledgements{
G.W. acknowledges Prof. Wei-Tou Ni for helpful comments. 
This work is supported by the National Key Research and Development Program of China under Grant No. 2021YFC2201903, and NSFC No. 12003059.
This work are performed by using the python packages \textsf{numpy} \cite{harris2020array}, \textsf{scipy} \cite{2020SciPy-NMeth}, \textsf{pandas} \cite{pandas}, \textsf{MultiNest} \cite{Feroz:2008xx} and \textsf{PyMultiNest} \cite{Buchner:2014nha}, and the plots are make by utilizing $\mathsf{matplotlib}$ \cite{Hunter:2007ouj}, \textsf{GetDist} \cite{Lewis:2019xzd}. The GW signals are generated using \textsf{LALSuite} \cite{lalsuite,swiglal} and \textsf{PyCBC} \cite{alex_nitz_2024_10473621}.
\\ \\
\textbf{Conflict of Interest} The authors declare that they have no conflict of interest.
}

\nocite{*}
\bibliographystyle{scpma}
\bibliography{refs}

\begin{appendix}

\renewcommand{\thesection}{Appendix}
\section{}

\subsection{Orthogonalization of TDI observables} \label{sec:orthogonalizatble}

\begin{figure}[H]
\centering
\includegraphics[width=0.48\textwidth]{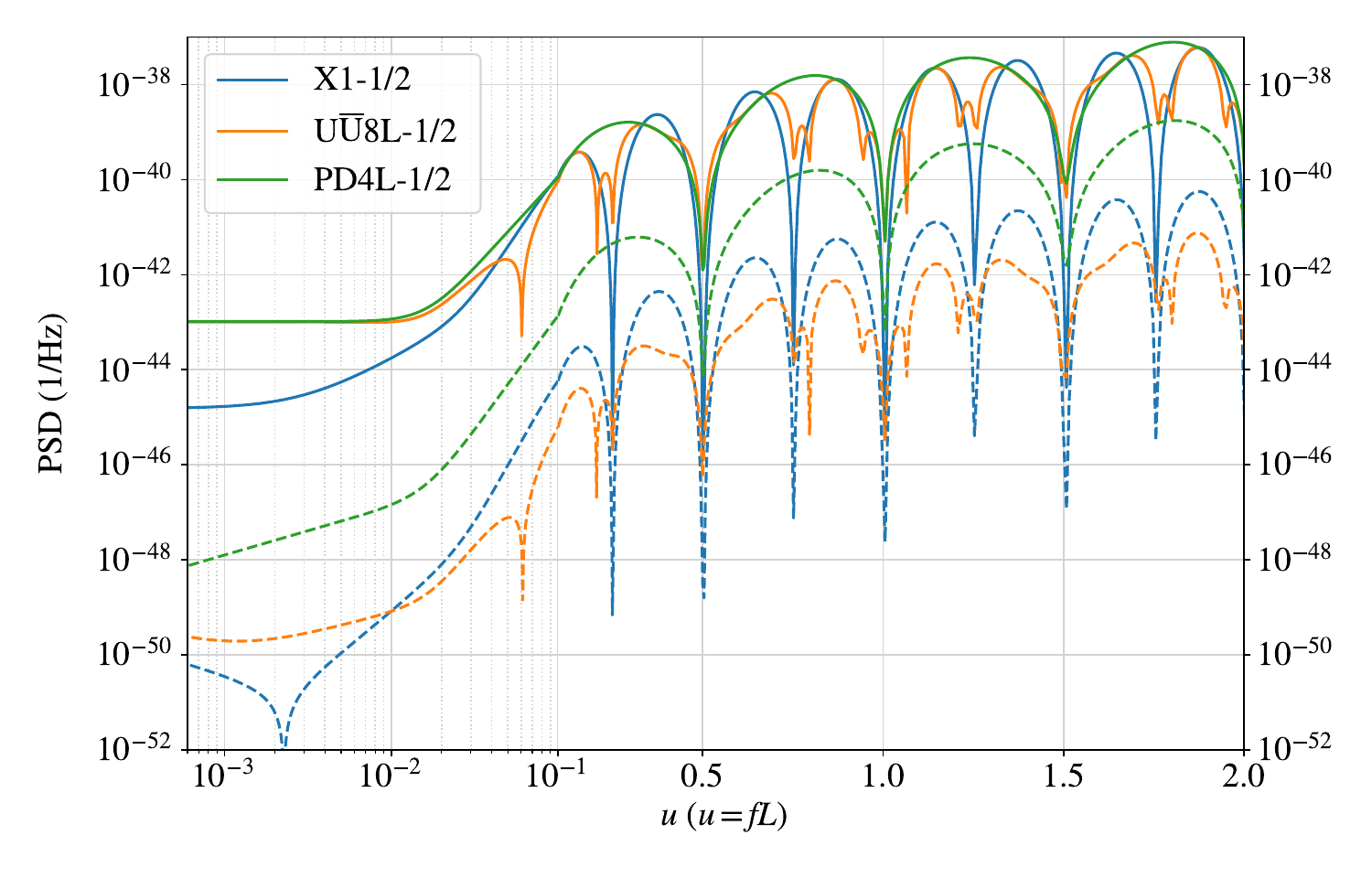}
\caption{\label{fig:TDI_csd_real_imag} The real (solid curves) and imaginary (dashed curves) components of the CSDs between two ordinary TDI channels for selected scheme. The blue, orange, and green curves represent the CSDs from Michelson, hybrid Relay and PD4L, respectively. Each pair of solid and dashed curves represents the respective real and imaginary parts of a CSD. The real components are several orders of magnitude larger than the imaginary parts, ensuing the validity of the transformation from ordinary channels to (quasi-)orthogonal channels. }
\end{figure}

The noise covariance matrix from three ordinary channels $(a, b, c)$ of a TDI configuration can be expressed as:
\begin{equation}
\begin{bmatrix}
S_\mathrm{a} & S_\mathrm{ab} & S_\mathrm{ac} \\
S_\mathrm{ba} & S_\mathrm{b} & S_\mathrm{bc} \\
S_\mathrm{ca} & S_\mathrm{cb} & S_\mathrm{c}
\end{bmatrix}
\simeq 
\begin{bmatrix}
S_\mathrm{a} & S_\mathrm{ab} & S_\mathrm{ab} \\
S_\mathrm{ab} & S_\mathrm{a} & S_\mathrm{ab} \\
S_\mathrm{ab} & S_\mathrm{ab} & S_\mathrm{a}
\end{bmatrix}, 
\end{equation}
where $S_\mathrm{a}$ represents PSD of data stream $a$, and $S_\mathrm{ab}$ denotes CSD between streams $a$ and $b$.
The CSDs of $S_{ab}$ and $S_{ba}$ are complex conjugates. However, the orthogonal transform remains a good approximation if two conditions are met: 1) the real components of three CSD pairs are approximately equal, a condition readily met due to constellation symmetry, and 2) the real parts of the CSDs significantly dominate the imaginary parts.
Figure \ref{fig:TDI_csd_real_imag} shows the real (solid) and imaginary (dashed) parts of the CSDs between first two ordinary channels in the Michelson, hybrid Relay, and PD4L configurations. Across all three configurations, the real components are several orders of magnitude larger than their imaginary components, supporting the validity of the (quasi-)orthogonal transformation described in Eq. \eqref{eq:abc2AET}.

\subsection{laser noise and clock noise suppression} \label{sec:noise_suppression}

\begin{figure}[H]
\centering
\includegraphics[width=0.45\textwidth]{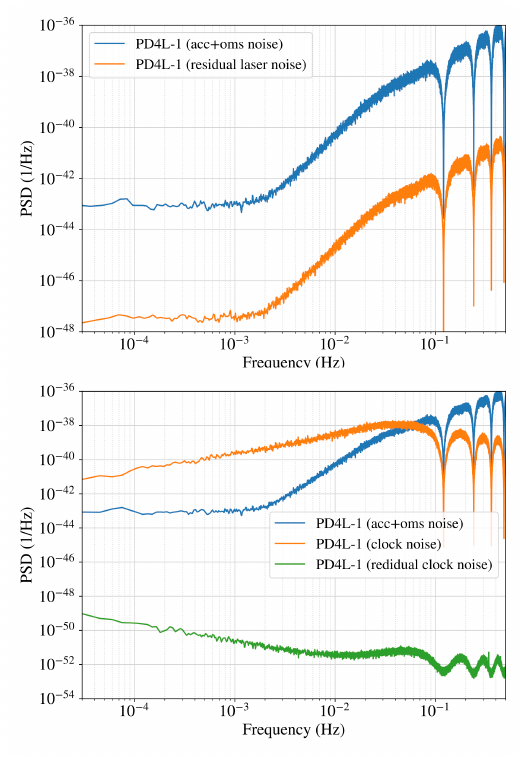}
\caption{\label{fig:noise_suppression} Spectrum of residual laser noise (upper panel) and clock noise (lower panel) for the PD4L-1 channel. Secondary noise, including acceleration noise and OMS noise, are shown in blue for comparison. In the upper plot, residual laser noise is several orders of magnitude lower than the secondary noise. In the lower plot, the clock jitter noise (orange curve) is higher than secondary noise in low frequencies, while the residual clock noise (green curve), after applying the subtraction method from \cite{Hartwig:2020tdu}, is significantly below the secondary noise level. }
\end{figure}

As a second-generation TDI configuration, it is essential for PD4L to sufficiently suppress the laser frequency noise. Assuming a laser stability of $30 \ \mathrm{Hz}/\sqrt{\mathrm{Hz}}$, this corresponds to a noise PSD of $1 \times 10^{-26}$/Hz.
The ranging error between S/C is modeled with a PSD of $10^{-15} \frac{\mathrm{Hz}}{f} \ \mathrm{s/\sqrt{Hz}}$ and an additional bias of 3 ns (0.9 m) \cite{Bayle:2022okx,Staab:2023qrb}. Based on these noise assumptions, the residual laser noise in PD4L-1 is evaluated and shown by the orange curves in the upper panel of Figure \ref{fig:noise_suppression}. For comparison, secondary noises sources, including acceleration noise and OMS noise, are shown as blue curve. As seen in the plot, the residual laser noise in PD4L is several orders of magnitude below the secondary noise level.

Clock noise in PD4L-1 is assessed assuming a clock instability of $4 \times 10^{-27} / f $ in fractional frequency deviations. As illustrated in the lower panel of Figure \ref{fig:noise_suppression}, the clock noise in TDI observable exceeds secondary noise at lower frequencies. However, after applying the subtraction proposed in \cite{Hartwig:2020tdu}, the residual clock noise (green curve) is reduced to well below the secondary noise level.
These results confirm that the PD4L scheme meet the fundamental requirement for effective suppression of both laser and clock noise.

\subsection{Inference with mixed TDI combination set} \label{sec:hybrid_relay_zeta}

\begin{figure*}[htb]
\centering
\includegraphics[width=0.8\textwidth]{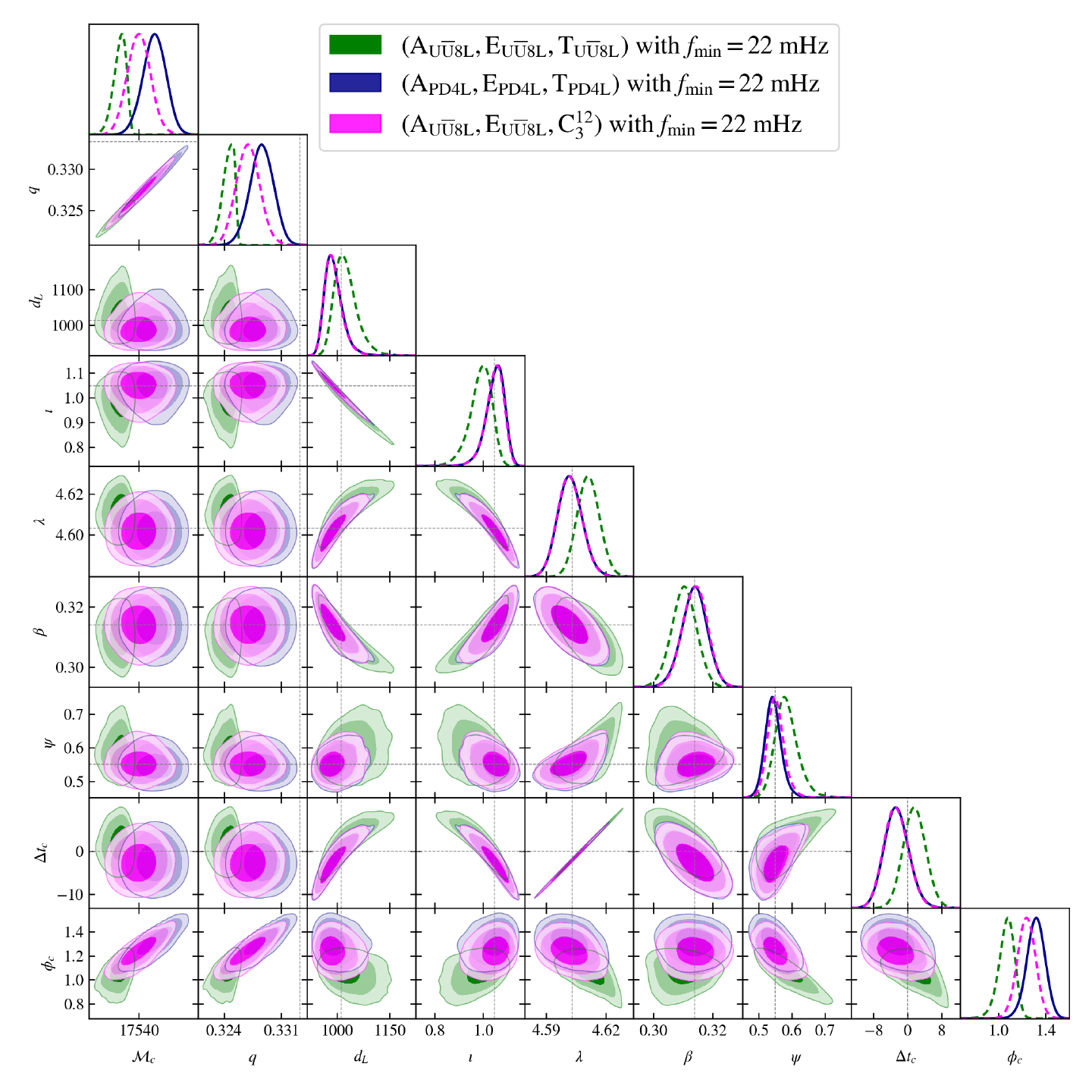}
\caption{\label{fig:corner_mbbh_mix} Inferred parameter distributions from three TDI schemes for a coalescing binary $(m_1=3 \times 10^4, m_2 = 10^4) \mathrm{M}_\odot$ at redshift $z=0.2$. The frequency range is set to [22 mHz, 0.4 Hz]. The green and blue regions respectively show the distribution obtained from optimal channels of hybrid Relay and PD4L, which shown previously in Figure \ref{fig:corner_noises_PD}. The magenta regions indicate result using the mixed observable set (A$_\mathrm{U\overline{U}8L}$, E$_\mathrm{U\overline{U}8L}$, $C^{12}_3$), which combines the science channels of hybrid Relay and the null stream $C^{12}_3$. The contours represent $1\sigma$, $2\sigma$, and $3\sigma$ confidence regions from darker to lighter shading. The grey dashed lines mark the injected (true) parameter values.
}
\end{figure*}

An optimal TDI combination, (A$_\mathrm{U\overline{U}8L}$, E$_\mathrm{U\overline{U}8L}$, $C^{12}_3$), was proposed in \cite{Wang:2024hgv} to enhance noise characterization. To assess its performance in analyzing signal in the high frequency band, we perform parameter inference using this combination, with the results shown by magenta colors in Figure \ref{fig:corner_mbbh_mix}. Compared to the results from hybrid Relay (green colors) and PD4L (blue colors), the parameter distributions (magenta) from (A$_\mathrm{U\overline{U}8L}$, E$_\mathrm{U\overline{U}8L}$, $C^{12}_3$) largely overlap with the result obtained from PD4L (A$_\mathrm{PD4L}$, E$_\mathrm{PD4L}$, T$_\mathrm{PD4L}$), except for two mass parameters. We infer that the $C^{12}_3$ channel helps reduce the null frequencies present in T$_\mathrm{U\overline{U}8L}$, resulting in distributions for extrinsic parameters that is consistent with that of PD4L. For the intrinsic parameter, the $C^{12}_2$ with $2L$ time span, could reduce frequency aliasing in the TDI process. However, its performance still falls short of PD4L, indicating that PD4L offer superior capability in analyzing the signal in the high-frequency regime.

\end{appendix}

\end{multicols}
\end{document}